\documentclass[final,3p,times,twocolumn]{elsarticle}
\usepackage[product-units=bracket-power, range-phrase=--, range-units=single, list-units=single, list-exponents=combine, separate-uncertainty]{siunitx}
\usepackage{subcaption}
\usepackage{wrapfig}
\usepackage{hyperref}
\usepackage{todonotes}
\usepackage{upgreek}
\usepackage{lineno}

\DeclareSIUnit\neq{\mathrm{n}_{\mathrm{eq}}\,\mathrm{cm^{-2}}}
\DeclareSIUnit\bit{\mathrm{-bit}}
\DeclareSIUnit\electrons{\mathrm{e}^-}
\DeclareSIUnit\electronvolt{\mathrm{eV}}
\DeclareSIUnit\rad{\mathrm{rad}}
\DeclareSIUnit\vcal{\Delta\mathrm{VCAL}}

\def\figurewidth{\linewidth}

\begin{document}

    \begin{frontmatter}
        \title{Cross talk of a large-scale depleted monolithic active pixel sensor (DMAPS) in \qty{180}{\nano\meter} CMOS technology}
        
        \author[1]{Lars Schall\corref{cor1}}
        \ead{lars.schall@uni-bonn.de}
        \author[1]{Christian Bespin}
        \author[1]{Ivan Caicedo}
        \author[1]{Jochen Dingfelder}
        \author[1]{Tomasz Hemperek\fnref{fn1}}
        \author[1]{Toko Hirono\fnref{fn2}}
        \author[1]{Fabian Hügging}
        \author[1]{Hans Krüger}
        \author[1]{Konstantinos Moustakas\fnref{fn3}}
        \author[5]{Heinz Pernegger}
        \author[5]{Petra Riedler}
        \author[5]{Walter Snoeys}
        \author[1]{Norbert Wermes}
        \author[1]{Sinuo Zhang}
        \cortext[cor1]{Corresponding author}
        \fntext[fn1]{Now at: DECTRIS AG, Baden-Dättwil, Switzerland}
        \fntext[fn2]{Now at: Karlsruher Insitut für Technologie, Karlsruhe, Germany}
        \fntext[fn3]{Now at: Paul Scherrer Institut, Villingen, Switzerland}
    
        \affiliation[1]{organization={Physikalisches Institut der Universität Bonn},
                    addressline={Nußallee 12}, 
                    city={Bonn},
                    country={Germany}}
  
        \affiliation[5]{organization={CERN},
                        addressline={Espl. des Particules 1},
                        city={Meyrin},
                        country={Switzerland}}
    
        \begin{abstract}
            Monolithic pixel detectors combine readout electronics and sensor in a single entity of silicon, which simplifies the production procedure and lowers the material budget compared to conventional hybrid pixel detector concepts.
            Benefiting from the advances in commercial CMOS processes towards large biasing voltage capabilities and the increasing availability of high-resistivity substrates, depleted monolithic active pixel sensors (DMAPS) are able to cope with the high-rate and high-radiation environments faced in modern high-energy physics experiments.
            TJ-Monopix2 is the latest iteration of a DMAPS development line designed in \qty{180}{\nano\meter} TowerSemicondutor technology, which features a large scale \qtyproduct{2x2}{\centi\meter} chip divided into \qtyproduct{512x512}{\mathrm{pixels}} with a pitch of \qtyproduct{33x33}{\micro\meter}.
            All in-pixel electronics are separated from its small collection electrode and process modifications are implemented to improve charge collection efficiency especially after irradiation.
            The latest laboratory measurements and investigations of a threshold variation observed for TJ-Monopix2 in typical operating conditions are presented.
        \end{abstract}
        
    \end{frontmatter}


\section{Introduction}
The recent advancements in commercial CMOS technologies propel the development of monolithic active pixel sensors (MAPS), which combine readout electronics and  sensor in a single piece of silicon~\cite{Turchetta:2001dy}.
This design approach simplifies the production procedure and reduces the material budget compared to the hybrid pixel detector concept.
Additional use of high-resistivity bulk substrates ($\uprho > \qty{1}{\kilo\ohm\cdot\centi\meter}$) combined with sufficiently large bias voltage capabilities facilitate the depletion of the sensitive volume and improve the fast charge collection by drift across a pixel~\cite{Peric:2007zz, Barbero:2019bkw}.
Hence, depleted MAPS (DMAPS) have an increased radiation tolerance making them a promising candidate for high-rate and high-radiation environment applications as faced in modern high-energy particle physics experiments.
By implementing a small collection electrode relative to the pixel pitch, the in-pixel electronics need to be separated and are placed in distinct p-wells. 
This design approach yields a reduction in sensor capacitance, thereby facilitating both low power consumption and enhanced threshold performance.
The long drift distances to the small collection node render the sensor more prone to radiation damage that require additional process modifications in the sensor layout.
The latest prototype of the TJ-Monopix DMAPS development line, TJ-Monopix2, is a large-scale chip designed in \qty{180}{\nano\meter} CMOS technology and high-resistivity substrate for compliance with the outer layer requirements of the ATLAS Inner Tracker upgrade~\cite{ATLAS:2017, Dingfelder:VERTEX2021}.
The observation of a threshold variation for TJ-Monopix2 lead to an examination trying to identify the cause, magnitude, and possible fixes, which are presented in this contribution.

\section{TJ-Monopix2}

\begin{figure}[htb]
    \centering
    \includegraphics[ width=\figurewidth]{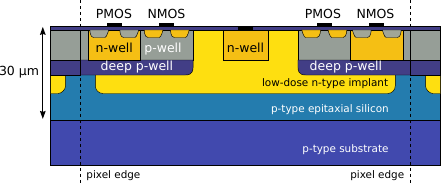}
    \caption{Schematic cross-section of a single TJ-Monopix2 pixel cell.
    The low-dose n-implant below the collection electrode facilitates a uniform electric field across the sensor while the gaps in the n-implant layer towards the edges improve the field shaping in the corners of a pixel.}
    \label{fig:crosssection}
\end{figure}

TJ-Monopix2 is a full scale \qtyproduct{2x2}{\centi\meter} DMAPS prototype designed in \qty{180}{\nano\meter} TowerSemiconductor\footnote{\url{https://towersemi.com/}} CMOS technology.
Based on the ALPIDE pixel detector~\cite{Mager:2016yvj}, its small collection electrode relative to the pixel pitch of \qtyproduct{33.04x33.04}{\micro\meter} is separated from the in-pixel electronics, which are housed in p-wells at the pixel edges.
The reduced sensor capacitance of $\mathcal{O}(\qty{3}{\femto\farad})$ accomplished by this design approach renders low analog power consumption of \qty{\sim1}{\micro\watt} and minimal noise operation at \qty{\sim5}{\electrons}.
The extended drift distances to the collection electrode and possible regions with low electric field expose the sensor to an increased probability of charge trapping after irradiation, making the sensor more susceptible to radiation damage.
To reach the design target in radiation tolerance of \qty{1e15}{\neq} fluence, a low-dose n-implant beneath the collection electrode is implemented to ensure the uniformity of the electric field across the entire sensor~\cite{Snoeys:2017hjn, Pernegger:2021cug}.
Previous studies have shown that an additional modification in form of a gap in the n-implant is needed to improve the lateral field shaping below the pixel corners after irradiation~\cite{Caicedo:2019lrk, Bespin:2020hge, Dyndal:2019nxt}.
Figure~\ref{fig:crosssection} shows a schematic cross-section of a single pixel cell.

TJ-Monopix2 is equipped with the synchronous column-drain readout architecture developed for the FE-I3 readout chip~\cite{Peric:2006km}.
A \qty{7}{\bit} gray encoded \qty{40}{\mega\hertz} counter (in the following referred to as \texttt{BCID} counter) is distributed along each double column.
Upon a registered hit, the leading and trailing edge (LE/TE) of a discriminated signal is sampled on a pixel level for charge measurements via the time-over-threshold method.
The corresponding \qty{40}{\mega\hertz} clock is internally derived from the \qty{160}{\mega\hertz} command clock in the chip's periphery.
Furthermore, each pixel contains a \qty{3}{\bit} local threshold DAC to minimize the threshold dispersion across the chip.
A digital in-pixel injection circuitry is implemented to verify the general functionality of the analog front-end and readout architecture.

\section{Cross Talk Analysis}

\begin{figure}[htb]
    \centering
    \includegraphics[width=\figurewidth]{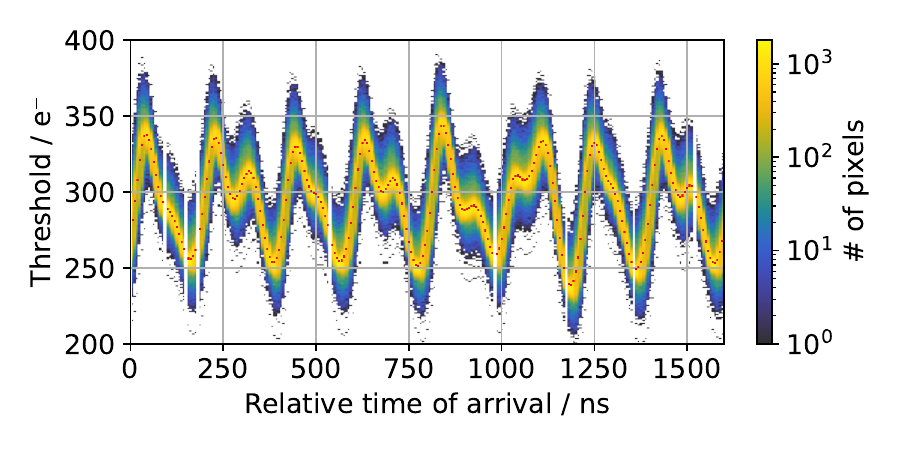}
    \caption{Tuned threshold distribution for $\mathcal{O}(\qty{30000}{pixels})$ measured at different relative hit arrival times with respect to the \texttt{BCID} counter running at $\qty{40}{\mega\hertz}$.
    The section shown corresponds to half of the \qty{7}{\bit} full counter resolution.
    The red dots indicate the average threshold over all activated pixels at each measurement point.
    A periodic pattern of the threshold response relative to the arrival time of the hits is visible.
    }
    \label{fig:THRoscillation}
\end{figure}

\begin{figure}[htb]
    \centering
    \includegraphics[width=\figurewidth]{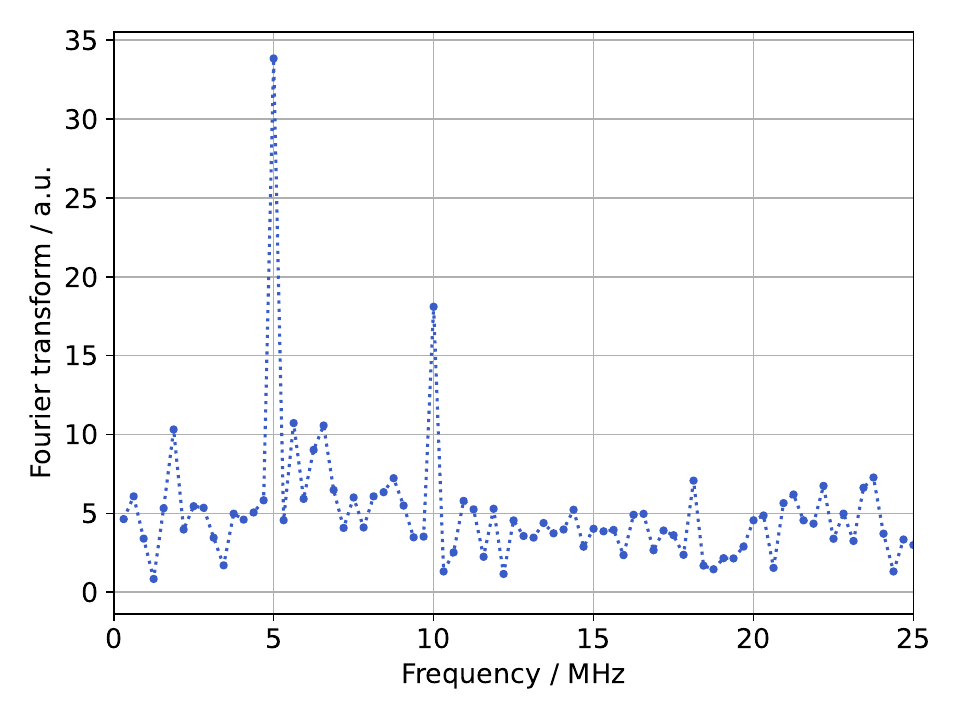}
    \caption{Fourier analysis of the periodic threshold variation shown in Figure~\ref{fig:THRoscillation}.
    The analysis reveals a dominant frequency of \qty{5}{\mega\hertz} and a less prominent \qty{10}{\mega\hertz} component.}
    \label{fig:FFT}
\end{figure}

Extensive tests of TJ-Monopix2 have verified improvements in threshold and equivalent noise charge (ENC) performance to its predecessor, TJ-Monopix1, in laboratory conditions~\cite{Bespin:2023vyw}.
Throughout the characterization, variations in the threshold response of TJ-Monopix2 were observed.
To investigate this, the threshold was measured relative to a fixed time of arrival for all hits.
The time of injection with respect to the \texttt{BCID} counter was shifted in steps of \qty{3.125}{\nano\second} by resetting the counter and adding an adjustable delay before every injection.
A periodic pattern of the threshold distribution depending on the hit arrival time was observed, which is displayed in Figure~\ref{fig:THRoscillation} over an exemplary $64\cdot\qty{25}{\nano\second} = \qty{1600}{\nano\second}$ interval of the \qty{7}{\bit} counter.
Red markers visualize the average threshold for each measurement point.
The measurement was conducted for $\mathcal{O}(\qty{30000}{pixels})$ initially tuned to a mean threshold of \qty{260}{\electrons} with a dispersion of \qty{6}{\electrons}.\footnote{A conversion factor of \qty{8.8}{\electrons} per injected charge DAC unit was applied corresponding to a calibration measurement utilizing Fe55 for this chip.}
Within the full \texttt{BCID} counter cycle, a maximum variation in mean threshold of about \qty{105}{\electrons} was quantified, which exceeds the very good ENC performance of \qty{7}{\electrons} by a factor of \qty{15}{}.
Figure~\ref{fig:FFT} shows a Fourier analysis of the periodic threshold pattern revealing a dominant frequency of \qty{5}{\mega\hertz} and a lesser noticeable \qty{10}{\mega\hertz} component.
Due to the gray encoding used in the \texttt{BCID} counter, the observed frequencies are equal to the toggling frequency of the two least significant bits of the counter.
Table~\ref{tab:graycode} demonstrates the gray encoded counting method employing the first eight clock cycles, thereby illustrating the resulting toggling frequencies of \qty{10}{\mega\hertz} and \qty{5}{\mega\hertz}.
\begin{table}[htb]
    \centering
        \begin{tabular}{|c|c|c|}\hline
             & \multicolumn{2}{|c|}{LE/TE counter}\\\hline
            Time & Decimal & Gray encoded \\\hline
            \qty{0}{\nano\second} & $0$ & $0000$ \\\hline
            \qty{25}{\nano\second} & $1$ & $0001$\\\hline
            \qty{50}{\nano\second} & $2$ & $0011$\\\hline
            \qty{75}{\nano\second} & $3$ & $0010$\\\hline
            \qty{100}{\nano\second} & $4$ & $0110$\\\hline
            \qty{125}{\nano\second} & $5$ & $0111$\\\hline
            \qty{150}{\nano\second} & $6$ & $0101$\\\hline
            \qty{175}{\nano\second} & $7$ & $0100$\\\hline
        \end{tabular}
    \caption{Exemplary gray encoding of the first eight \texttt{BCID} counter values.
    Given the default \qty{40}{\mega\hertz} clock displayed in the first column, the least significant bit toggles at \qty{10}{\mega\hertz}.
    The toggling frequency is halved for each additional bit.
    }
    \label{tab:graycode}
\end{table}

\begin{figure}[htb]
    \centering
    \includegraphics[width=\figurewidth]{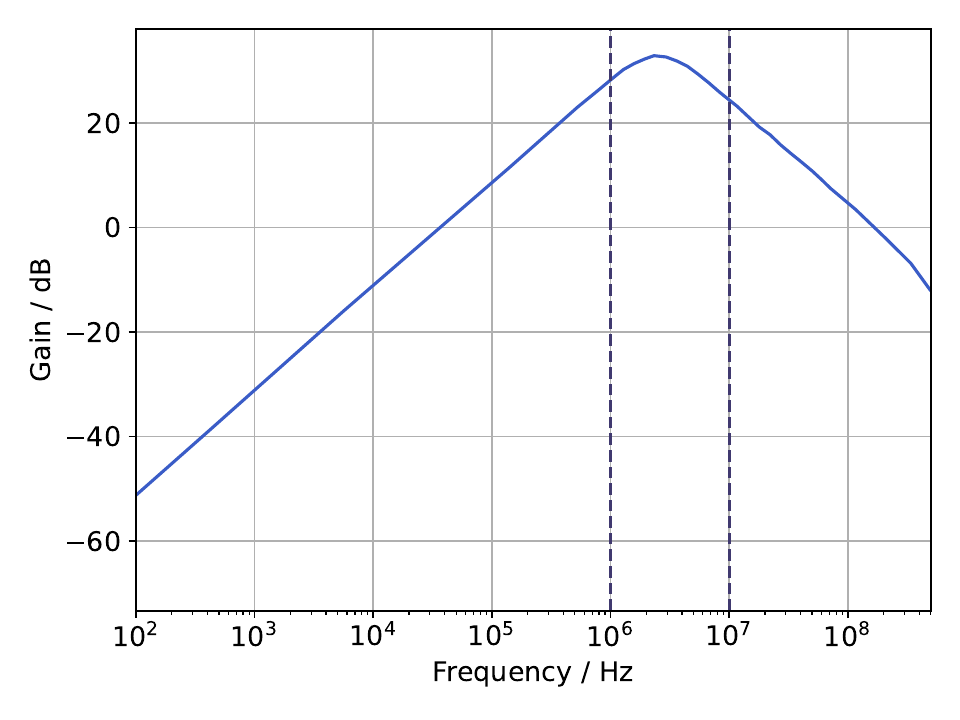}
    \caption{Simulated transfer function of the pre-amplifier implemented in TJ-Monopix2 (modified from~\cite{Moustakas:2021}).
    The dashed lines indicate the range between \qtyrange{1}{10}{\mega\hertz} in which the maximum amplification is reached.
    The exact frequency of the highest gain depends on the exact input capacitance, biasing, and feedback current settings.}
    \label{fig:transferfunction}
\end{figure}

The simulated transfer function of the pre-amplifier implemented in TJ-Monopix2 is shown in Figure~\ref{fig:transferfunction}.
The highest amplification is achieved for signals in the frequency range between \qtyrange{1}{10}{\mega\hertz}~\cite{Moustakas:2021}.
Thus, the pre-amplifier facilitates potential cross-talk precisely in the range of frequencies corresponding to the toggling of the \texttt{BCID} counter bits.

\begin{figure}[htb]
    \centering
    \includegraphics[width=\figurewidth]{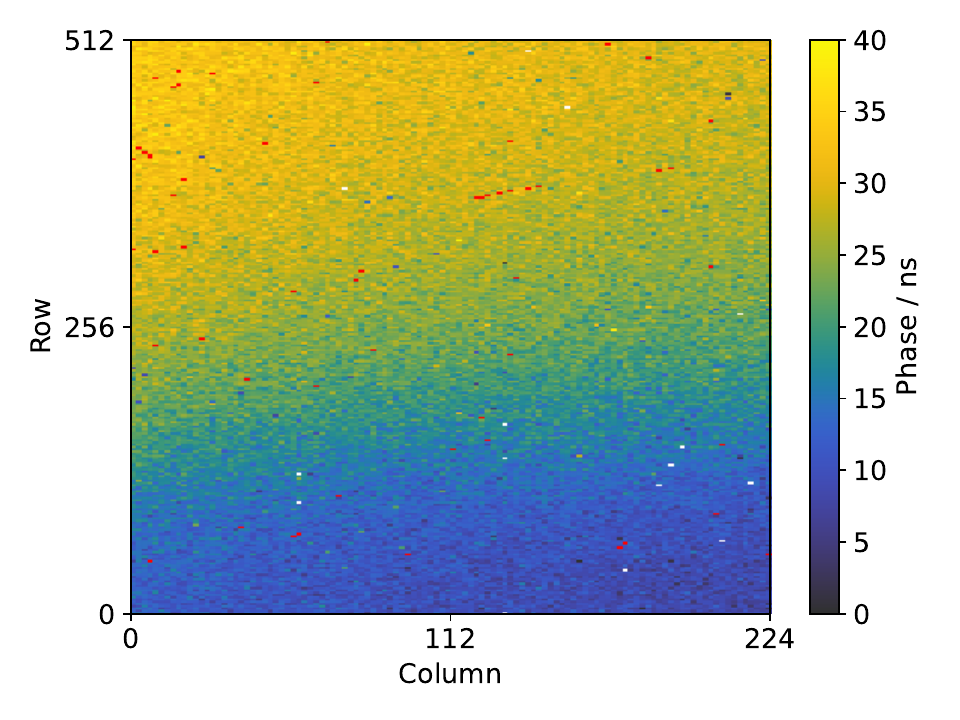}
    \caption{Per pixel phase estimation of the threshold variation shown in Figure~\ref{fig:THRoscillation}.
    Only every fourth pixel in this map was enabled to optimize scan time.
    The gradient from bottom to top originates from the counter and injection pulse propagation time across the matrix resulting in a local dependency of the threshold variation.}
    \label{fig:phasemap}
\end{figure}

\begin{figure}[htb]
    \centering
    \includegraphics[width=\figurewidth]{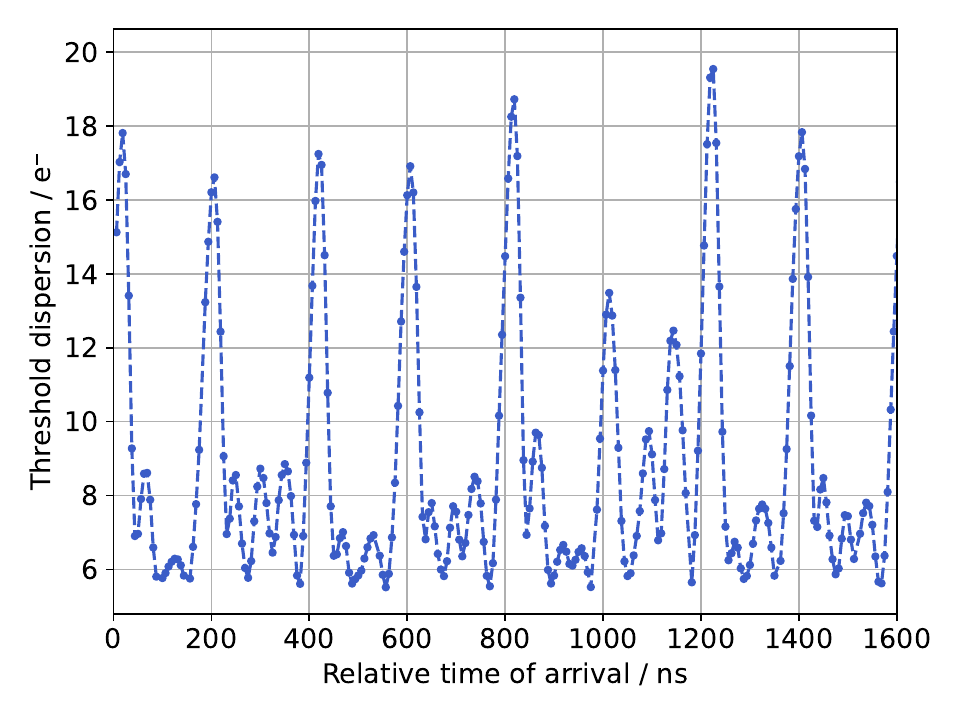}
    \caption{Threshold dispersion relative to the hit arrival times with respect to the LE counter.
    The periodic spikes up to \qty{20}{\electrons} are a direct consequence of the clock propagation time and the resulting phase difference in the threshold variation.}
    \label{fig:THRdispersion}
\end{figure}

Furthermore, a delay in the threshold variation based on pixel position within the matrix was observed, arising from the influence of the \texttt{BCID} counter and injection pulse propagation time.
To get an estimation of the local dependency, a simple sine function was fitted to the respective periodic pattern in threshold response for each pixel.
The estimated phase of the threshold variation across the matrix is displayed in Figure~\ref{fig:phasemap}, while frequency and amplitude of the variation remained similar for all pixels.
Since the chip periphery and the end of column logic is situated at the bottom of the chip, the observed top to bottom gradient in the phase can be explained.
This gradient in the phase leads to variations in the threshold dispersion relative to the arrival time of hits.
Figure~\ref{fig:THRdispersion} shows the evolution of the threshold dispersion over half of the maximum \texttt{BCID} counter interval.
Originating from the combination of threshold variation frequency and phase shift, a periodic pattern in the threshold dispersion is visible.
The dispersion reaches up to \qty{20}{\electrons} in case of the most adverse phase difference between top and bottom pixels, surpassing the initial tuning result of \qty{6}{\electrons} by more than threefold.

\begin{figure}[htb]
    \centering
    \includegraphics[width=\figurewidth]{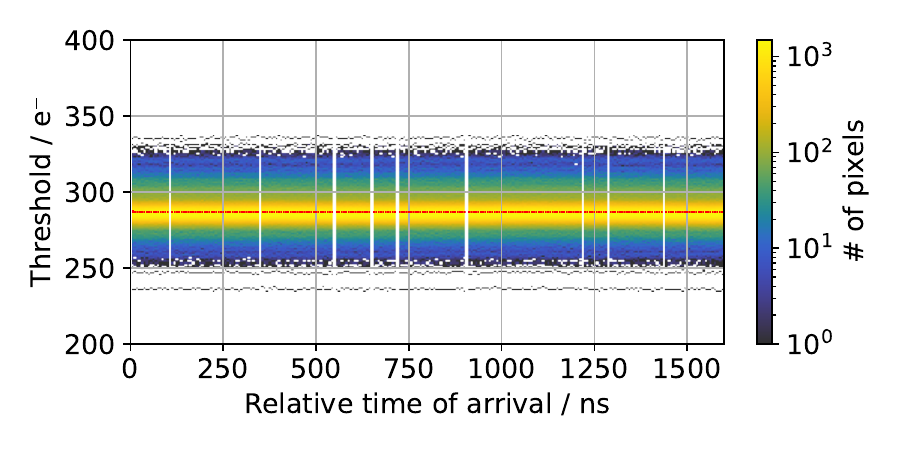}
    \caption{Variation of the tuned threshold distribution relative to the arrival time of hits for $\mathcal{O}(\qty{30000}{pixels})$.
    Disabling the LE/TE counter and the corresponding clock reduces the variation and maximum difference in average threshold across all pixels significantly.
    The red markers indicate the mean threshold of the activated pixels.}
    \label{fig:THRbcidoff}
\end{figure}

Disabling the distribution of the \texttt{BCID} counter across the matrix mitigated the threshold variation completely as shown in Figure~\ref{fig:THRbcidoff}.
In the absence of leading edge (and charge) information for hits, the identical assignment of measurement points to a specific time of arrival as previously calibrated is employed.
This proves the origin of the cross talk to arise from the distribution of the toggling \texttt{BCID} counter bits across the matrix.

\begin{figure}[htb]
    \centering
    \includegraphics[width=\figurewidth]{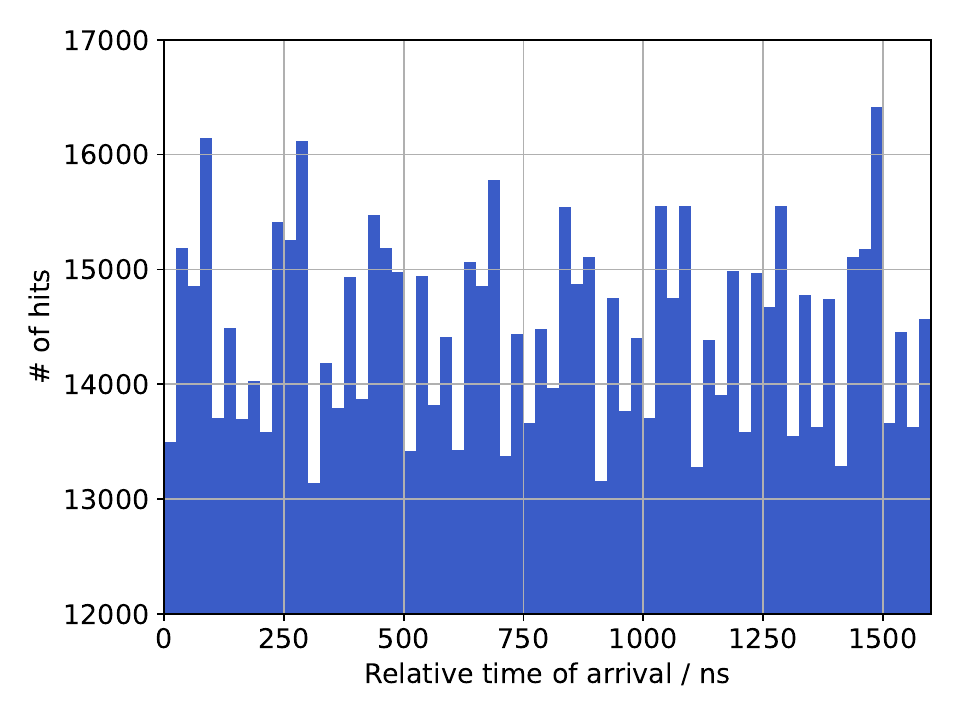}
    \caption{Relative hit arrival time of a measurement with a radioactive source for an exemplary section of the \texttt{BCID} counter.
    The periodic pattern analog to Figure~\ref{fig:THRoscillation} in relative arrival time of non-injected hits proves that the cross talk is affecting either the analog signal or the threshold of the chip and not the injection circuitry.}
    \label{fig:LE_distribution}
\end{figure}

To better understand which part of the chip is affected by this cross talk, a measurement with a radioactive source was analyzed for the relative time of arrival of all detected hits.
The resulting distribution over an exemplary segment of the complete \texttt{BCID} counter is shown in Figure~\ref{fig:LE_distribution}.
Analog to the observed threshold variation (Figure~\ref{fig:THRoscillation}), a periodic pattern in the relative hit arrival time of non-injected hits was identified indicating an increased detection rate of hits at periods of lower thresholds.
Consequently, this cross talk has no influence on the digital injection circuitry, but must impact the measured analog signal of a hit or the threshold level of the chip.

In an attempt to suppress the cross talk, the internally derived \qty{40}{\mega\hertz} clock was modulated by adjusting the command clock frequency to change the toggling frequency of the \texttt{BCID} counter bits.
It was not possible to successfully suppress the threshold variation for a feasible command clock.
In case of the deceleration of the clocks by a factor of four (yielding toggling frequencies of \qty{2.5}{\mega\hertz} and lower), a peak threshold fluctuation of \qty{90}{\electrons} remained.
Fourier analysis of the resulting curves resembled the respective change in toggling frequency of the counter bits within the sensitive pre-amplifier frequency range.

Upon closer investigation of the chip's layout, the responsible distribution lanes of the \texttt{BCID} counter across the matrix were found to be well-shielded.
Simulations of the chip's matrix design are currently ongoing to better understand the origin of this phenomenon.
For laboratory measurements, a predetermined delay between the reset of the \texttt{BCID} counter and the injection of a hit is used as a practical solution.
However, this approach is not applicable for measurements where there is no control over the hit arrival time and energy information is wanted, such as radioactive source or beam tests.

\section{Conclusion}

The threshold performance of TJ-Monopix2 DMAPS prototypes was further investigated and a modulation of the threshold response relative to the hit arrival time was discovered.
A maximum change of \qty{105}{\electrons} in the periodic pattern of the mean threshold was quantified, exceeding the ENC performance of the chip by a factor of \qty{15}{}.
Variations in threshold dispersion reaching up to three times the initial value after tuning were measured relative to the hit arrival time and followed the same periodic behavior.
Execution of a Fourier analysis on the mean threshold oscillation revealed a dominant \qty{5}{\mega\hertz} frequency, matching the toggling frequency of the second least significant \texttt{BCID} counter bit.
Analysis of measurements with a radioactive source has shown higher hit detection rates at periods of lower threshold, indicating that the cross talk affects either the analog signal of a hit or the threshold level of the chip.
Mitigation of the threshold variation by disabling the \texttt{BCID} counter distribution across the matrix has proven that the cross talk is caused by the toggling of the counter bits.
The constant phase relation between injection strobe and \texttt{BCID} counter, which is typically used for laboratory measurements, hides the now observed cross talk phenomenon.
For a better understanding of this behavior, simulations of the chip matrix are still ongoing. 

\section{Acknowledgments}
This project has received funding from the Deutsche Forschungsgemeinschaft DFG (grant WE 976/4-1), the German Federal Ministry of Education and Research BMBF (grant 05H15PDCA9), and the European Union’s Horizon 2020 research and innovation programme under grant agreements no. 675587 (Maria Sklodowska-Curie ITN STREAM), 654168 (AIDA-2020), and 101004761 (AIDAinnova).

\bibliographystyle{elsarticle-num}
\bibliography{bibliography.bib}


\end{document}